\documentstyle[11pt,aaspp4,flushrt]{article} 
\input psfig
\begin{document}
\title{Strongly Polarized Optical Afterglows of Gamma-Ray Bursts} 
\author{Andrei Gruzinov}
\affil{Institute for Advanced Study, School of Natural Sciences, Princeton, NJ 08540}

\begin{abstract}
The optical afterglows of the gamma ray bursts can be strongly polarized, in principle up to tens of percents, if: (i) the afterglow is synchrotron radiation from an ultra-relativistic blast, (ii) the blast is beamed during the afterglow phase, i.e. the shock propagates within a narrow jet, (iii) we observe at the right time from the right viewing angle, (iv) magnetic fields parallel and perpendicular to the jet have different proper strengths.

\end{abstract}

\section{Introduction, observations}
Gamma-ray burst (GRB) afterglows were explained by synchrotron radiation from an ultra-relativistic blast wave (Paczynski \& Rhoads 1993, Katz 1994, Meszaros \& Rees 1997, Vietri 1997, Waxman 1997). A good way to test the synchrotron emission model is to measure the polarization (Loeb \& Perna 1998, Gruzinov \& Waxman 1999). 

Hjorth et al (1999) found an upper limit of 2.3\% on the linear polarization of the optical emission from GRB 990123. Covino et al (1999) have detected a 1.7\% linear polarization of the optical transient associated to GRB 990510. A few percent polarization relative to the stars in the field can be induced by dust along the line of site in the host galaxy. To be sure that the polarization is from the GRB afterglow, one needs to look for the time variability of the polarization signal (Gruzinov \& Waxman 1999).

While current observations are not in disagreement with the model of Gruzinov \& Waxman (1999), we would like to bring attention to the fact that a much stronger polarization of the optical afterglows, tens of percents, is theoretically possible. Polarization of more than a few percent would be a true signature of the synchrotron emission model. 

The strong polarization might be achieved if the afterglow is beamed, and the magnetic fields parallel and perpendicular to the jet have different strengths (\S 2, Medvedev \& Loeb 1999). Medvedev \& Loeb (1999) believe that the magnetic fields in the synchrotron emitting plasma are strictly parallel to the shock front. We do not think that this is the case (\S 3), but their idea that parallel and perpendicular magnetic fields can have different strength seems plausible, especially in the jet scenario.

\section{Strongly polarized afterglows}

Synchrotron emission is strongly polarized. The degree of polarization $\Pi _0$ depends on the distribution function and the frequency, a typical value is $\Pi _0=60\%$. In an unresolved source like a GRB afterglow, the polarization will cancel out if the magnetic field is fully mixed in the emitting plasma. If the symmetry is violated, the unresolved image will have a non-zero polarization $\Pi$. 

Loeb \& Perna (1998) and Gruzinov \& Waxman (1999) break the symmetry by small number statistics - if the number of coherent magnetic patches in the synchrotron emitting plasma is $N$, the measured polarization is $\Pi \sim \Pi _0  /\sqrt{N}$. Gruzinov \& Waxman (1999) then estimate $N$ for the radiation from the self-similar ultra-relativistic blast of Blandford \& McKee (1976). Medvedev \& Loeb (1999) rely on interstellar scintillations in the radio band to resolve the image.

Medvedev \& Loeb (1999) in their paper on radio afterglows note that another way to break the symmetry is to have a beamed blast wave in which magnetic fields are perpendicular to the jet. We do not think that magnetic fields are purely perpendicular, but it seems plausible that magnetic fields are not fully mixed, i.e., the magnetic fields parallel and perpendicular to the jet have significantly different averaged strengths. A highly polarized optical afterglow can result. 

For illustrative purposes, Fig. 1,  assume that (i) the opening angle of the jet is much smaller than the viewing angle $\theta$, (ii) $\theta \ll 1$, (iii) at the time of emission, the Lorenz factor of the emitting plasma is $\gamma =\theta ^{-1}$. Then the degree of polarization is 
\begin{equation}
\Pi = \Pi _0 {<B_{\parallel }^2>-0.5<B_{\perp }^2>\over <B_{\parallel }^2>+0.5<B_{\perp }^2>},
\end{equation}
where the magnetic field is measured in the plasma rest frame, and we have assumed full mixing in the plane parallel to the shock: $<B_x^2>=<B_y^2>=0.5<B_{\perp }^2>$. 

\begin{figure}[htb]
\plotone{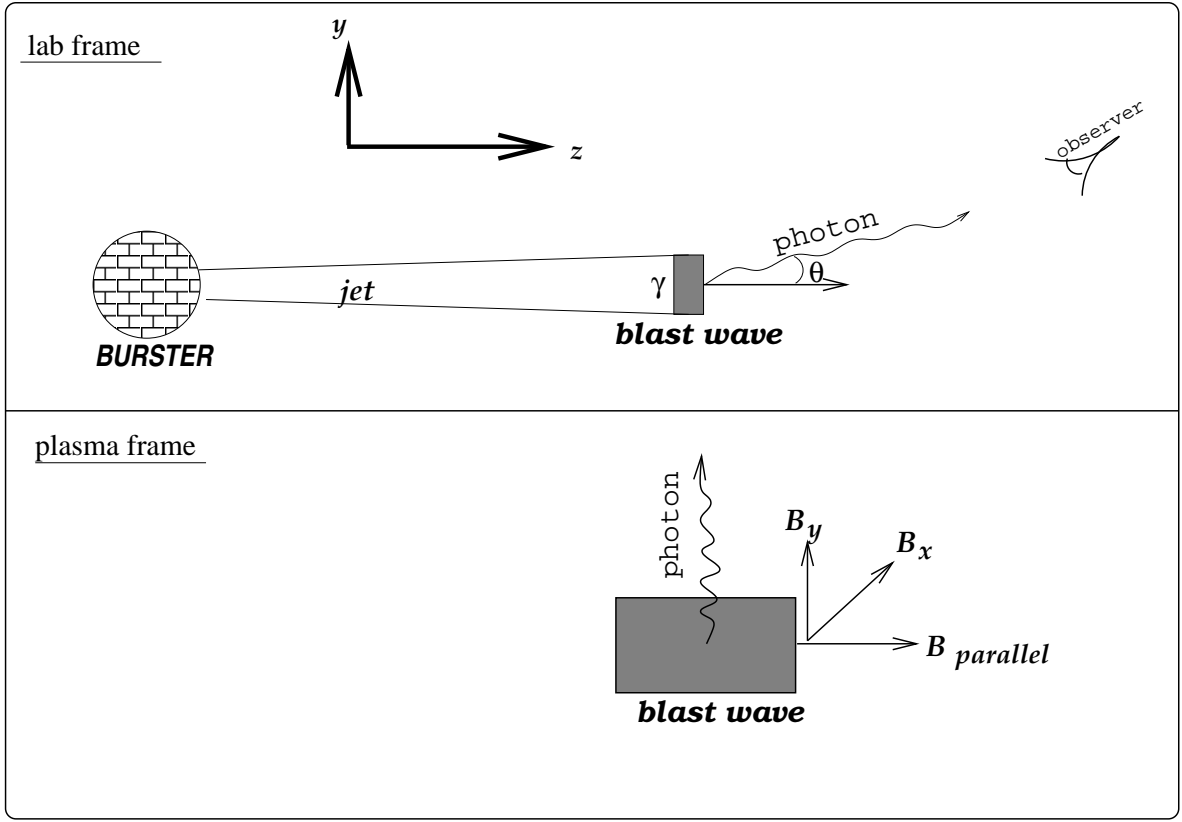}
\caption{Photon emission in the lab frame ($\gamma$-burster/observer frame) and in the plasma rest frame, obtained by a boost in the negative z direction with  Lorentz factor $\gamma$. The case of $\gamma =\theta ^{-1}$.}
\end{figure}

If $\theta \not= \gamma ^{-1}$, the photon makes an angle $\alpha \not= \pi /2$ with the z-axis in the plasma frame, and equation (1) should be replaced by
\begin{equation}
\Pi = \Pi _0 \sin ^2\alpha {<B_{\parallel }^2>-0.5<B_{\perp }^2>\over <B_{\parallel }^2>\sin ^2\alpha+0.5<B_{\perp }^2>(1+\cos ^2\alpha )},
\end{equation}
where 
\begin{equation}
\sin \alpha ={2\gamma \theta \over 1+\gamma ^2\theta ^2}.
\end{equation}
As long as the proper strengths of the fields are significantly different, and the viewing angle and the jet opening angle are $\sim \gamma ^{-1}$, the emission is strongly, $\sim \Pi _0 \sim$tens of percent, polarized. 

\section{The origin of the magnetic fields}
The origin of the magnetic fields remains the biggest problem of the blast synchrotron emission model. To explain the observed afterglows, the magnetic field in the shocked plasma has to have close to equipartition magnitude. The shock compression of the magnetic field of the surrounding medium is insufficient. In order to be in near equipartition after the passage of the ultra-relativistic shock, the magnetic fields in the unshocked medium must be in near equipartition with the rest mass energy density. The equipartition with the rest mass energy density in the cold unshocked medium is impossible - such fields could not be confined by the unshocked medium pressure. Therefore the unshocked medium is effectively unmagnetized, and magnetic fields must be generated by the blast wave itself.

As explained by Gruzinov \& Waxman (1999), the collisionless ultra-relativistic blast wave will generate strong magnetic fields by the Weibel instability (electromagnetic instability in a plasma with an anisotropic distribution function, e.g. Krall \& Trivelpiece 1973). These fields are generated at the microscopic, skin depth scales ($\delta \sim c/\omega _p$, $\omega _p$ is the plasma frequency). The skin depth is much smaller than the blast wave proper thickness $l$, typically $\delta /l\sim 10^{-10}$. Gruzinov \& Waxman (1999) suggested that the length scale of the magnetic field will grow after the shock transition by an unidentified mechanism. Medvedev \& Loeb (1999) believe that the small-scale Weibel-generated  fields solve the problem, but, unfortunately, these small scale fields cannot explain the magnetization of the bulk of the blast wave plasma.

The skin depth scales are exactly the dissipative scales. After a few inverse plasma frequency times, the Weibel instability should isotropise the plasma, bringing it up the shock adiabatic (Taub adiabatic). The small-scale fields should be damped. If Weibel instability were the full story, magnetic fields would have existed only in the narrow layer (few skins deep) in the vicinity of the shock front. Most of the blast wave plasma would be free of magnetic fields. The synchrotron emission would be negligible.

In reality, close to equipartition magnetic fields should exist on the large scales. This is what happens in the astrophysical plasmas where we can measure the magnetic fields. But we do not understand how the strong large-scale fields could be generated by the  blast wave. It may be that the unshocked plasma is already pre-heated by the jet, and the large-scale fields are close to equipartition even before the blast wave passage.

\section{Summary}

GRB optical afterglows might be strongly, up to tens of percents, polarized. We do not understand the origin of the magnetic field in the blast synchrotron emission model of the GRB afterglows, but observations of the highly polarized afterglows may demonstrate the presence of the magnetic field.

\acknowledgements I thank John Bahcall, Daniel Eisenstein, and David Hogg for useful discussions. I thank Bruce Draine and Pawan Kumar for useful information. This work was supported by NSF PHY-9513835.

\end{document}